\documentclass[conference]{IEEEtran}
\IEEEoverridecommandlockouts

\usepackage{cite}
\usepackage{hyperref}
\hypersetup{
    colorlinks= false,
    linkcolor=yellow,
    citecolor=yellow,
    urlcolor=black,
}
\usepackage{url}
\usepackage{amsmath,amssymb,amsfonts}
\usepackage{algorithmic}
\usepackage{graphicx}
\usepackage{textcomp}
\usepackage{longtable}
\usepackage{xcolor}
\usepackage{enumitem}
\usepackage{subcaption}
\usepackage{booktabs}
\def\BibTeX{{\rm B\kern-.05em{\sc i\kern-.025em b}\kern-.08em
    T\kern-.1667em\lower.7ex\hbox{E}\kern-.125emX}}

\renewcommand{\baselinestretch}{0.85}  
\usepackage[compatibility=false]{caption}
\begin{document}

\title{Seeing Heat with Color - RGB-Only Wildfire Temperature Inference from SAM-Guided Multimodal Distillation using Radiometric Ground Truth\\

\thanks{This material is based upon work supported by the National Aeronautics and Space Administration (NASA) under award number 80NSSC23K1393, the National Science Foundation under Grant Numbers CNS-2232048, and CNS-2204445.}

}

\author{\IEEEauthorblockN{Michael Marinaccio, Fatemeh Afghah}
\IEEEauthorblockA{Dept. of Electrical and Computer Engineering, Clemson University, Clemson, SC, USA \\
\{mmarina, fafghah\}@clemson.edu}

}

\maketitle 
\begin{abstract}
High-fidelity wildfire monitoring using Unmanned Aerial Vehicles (UAVs) typically requires multimodal sensing—especially RGB and thermal imagery—which increases hardware cost and power consumption. This paper introduces \textit{SAM-TIFF}, a novel teacher-student distillation framework for pixel-level wildfire temperature prediction and segmentation using RGB input only. A multimodal teacher network trained on paired RGB-Thermal imagery and radiometric TIFF ground truth distills knowledge to a unimodal RGB student network, enabling thermal-sensor-free inference. Segmentation supervision is generated using a hybrid approach of segment anything (SAM)-guided mask generation, and selection via TOPSIS, along with Canny edge detection and Otsu's thresholding pipeline for automatic point prompt selection.  Our method is the first to perform per-pixel temperature regression from RGB UAV data, demonstrating strong generalization on the recent FLAME 3 dataset. This work lays the foundation for lightweight, cost-effective UAV-based wildfire monitoring systems without thermal sensors.
\end{abstract}


\section{Introduction}

Wildfires' increasing scale, frequency, and destructiveness have underscored the urgent need for advanced situational awareness and rapid response systems. Unmanned Aerial Vehicles (UAVs) have become critical assets in this domain due to their ability to quickly reach inaccessible areas, capture high-resolution imagery, and support real-time decision-making without endangering human responders. Modern wildfire response efforts now rely heavily on UAVs to perform a wide range of tasks: detecting active flames in complex terrain, monitoring the spatial progression of fires, identifying residual hot zones after suppression efforts, and verifying that burn areas have cooled sufficiently to prevent reignition. Achieving high reliability in these tasks demands advanced vision-based models capable of performing precise fire segmentation and estimating spatial temperature distributions at the pixel level \cite{BOROUJENI2024102369}.

While conventional UAV-based fire detection systems typically depend on visible (RGB) and infrared (IR) imaging, there are significant limitations to both modalities. RGB imagery, though widely available and inexpensive, is often unreliable under challenging conditions such as smoke occlusion, night-time operations, or environments with low color contrast between flames and background. Conversely, thermal sensors provide direct temperature information and are effective in poor visibility, but are expensive, power-hungry, limiting flight time and scalability for large-scale deployments \cite{rajoli2024,flamefinder}. Furthermore, many wildfire tasks—such as determining the precise boundaries of active flame zones or estimating thermal gradients within a burn scar—require joint reasoning over texture, color, and temperature information, motivating the need for multimodal perception. However, practical constraints make it necessary to develop methods that can emulate the benefits of thermal sensing using RGB input alone.

To address these challenges, we propose \textit{SAM-TIFF}, a novel multimodal-to-unimodal knowledge distillation framework designed to enable high-fidelity wildfire monitoring using only RGB UAV imagery. Our system is motivated by the premise that while thermal information is essential for training high-performing models, it need not be available during inference if its utility can be distilled into a compact, RGB-only model. Specifically, we train a multimodal teacher network on paired RGB-Thermal images and radiometric thermal TIFF ground truth, and use it to supervise a student network that takes only RGB input. This student model is optimized to simultaneously segment fire regions and regress per-pixel temperature estimates, enabling a lightweight, thermal-free model suitable for deployment on UAV platforms with constrained payload and energy budgets.

Central to our framework is the use of the recently released \textit{FLAME 3} dataset \cite{flame_3_dataset,Flame3_paper}, which provides high-resolution UAV imagery from multiple prescribed burns, captured in both RGB and thermal modalities. Critically, FLAME 3 includes calibrated radiometric thermal TIFFs, offering pixel-wise temperature values in degrees Celsius—marking the first time such data has been made publicly available at scale for wildfire research. This dataset enables our model to perform real-valued temperature regression rather than relying on colormapped thermal images or binary fire/no-fire annotations. However, FLAME 3 does not include segmentation masks, presenting a major obstacle for supervised training. To overcome this, we develop a hybrid pseudo-labeling pipeline that combines traditional computer vision techniques with state-of-the-art foundation models: Canny edge detection, Otsu’s thresholding, and the Segment Anything Model (SAM) are integrated into a pipeline that automatically generates candidate fire masks. A multi-criteria decision-making algorithm based on TOPSIS is used to select the most accurate mask from among SAM's proposals, leveraging temperature patch means, temperature mean absolute differences, IoUs, and structural similarity metrics to ensure alignment with the radiometric ground truth.

This work introduces several novel contributions to the field of wildfire monitoring and aerial computer vision:

\begin{itemize}[leftmargin=*]
    \item We present the first system to perform {per-pixel wildfire temperature regression} using only RGB input, enabling cost-effective, thermal-free UAV deployments.
    \item We introduce a {multimodal teacher to unimodal student distillation architecture} specifically designed for both segmentation and temperature inference, extending knowledge distillation techniques to the wildfire domain.
    \item We leverage {radiometric thermal TIFFs} from the FLAME 3 dataset as true ground truth for temperature prediction—pioneering a new direction in temperature-aware UAV perception.
    \item We develop a robust, automated {Autopoint Locator, segmentation mask generation, and mask selection pipeline} using SAM, Canny edges, Otsu thresholding, and TOPSIS for high-quality pseudo-labels in the absence of manual annotation.
   
\end{itemize}

\section{UAV-Based Wildfire Monitoring Datasets}
While there are general fire datasets \cite{corsican,AIDER2,DINCER,DataClusterLabs} for classification, semantic segmentation, and object detection exist, finding UAV-collected, wildfire-specific datasets are difficult to find \cite{Firedata}. Arguably the most robust and notable wildfire specific datasets are from the FLAME dataset series, containing paired drone aerial imagery, with FLAME 1 \cite{flame_1_dataset} and FLAME 2 \cite{flame_2_dataset} for classification and segmentation tasks, and most recently, FLAME 3 \cite{flame_3_dataset} for classification, modeling, segmentation, and temperature prediction tasks. 

\subsection{FLAME 2 Dataset}
The FLAME 2 Dataset \cite{flame_2_dataset}, released in 2022, contains drone-collected, paired RGB-T imagery, from a prescribed burn in Northern Arizona, and supplemental burn data. FLAME 2 provides RGB-T paired video frames labeled Fire / No Fire and Smoke / No Smoke for classification tasks. Overlap in the labels for fire and smoke provides opportunity for more specific classification and training models to recognize the presence of fire even if no visible flames are in the image. The downside of FLAME 2 is the high temporal resolution and singular burn environment the footage was captured in, making generalization more difficult. 

\subsection{FLAME 3 Dataset}
In December 2024, the successor to FLAME 2 was released as FLAME 3, which expands the data collection procedures and addresses many of the issues seen from FLAME 2 \cite{flame_3_dataset}. In FLAME 3, paired drone-collected multispectral imagery is provided from six rural prescribed burn locations, alongside the release of a new data type, radiometric thermal TIFFs. FLAME 3 aims to improve generalization capability by providing imagery across different burn locations, providing better aligned paired modality data through FOV corrections and cropping procedures, and inspiring a new generation of wildfire temperature inference models through the use of the TIFF data as ground truth. The TIFFs in FLAME 3 are represented as a 2D data structure containing raw temperature values in degrees Celsius, which is more machine-interpretable and different from using three-channel thermal colormapped images. With temperature-labeled pixels, this enables a wide range of applications for wildfire hotspot identification and more detailed fire region understanding, and is something that had not yet been seen until this dataset release. 

\subsection{Thermal TIFF Analysis and Calibration}

The TIFF data across burn locations in FLAME 3 contains varying temperature values. This is due to camera settings, weather, fuel type, burn intensity, etc., which impact temperature values recorded. Therefore, an approach top unify each burn location's TIFF data is needed. Based on information provided in the FLAME 3 \cite{flame_3_dataset} paper, we expect the Willamette Valley burn to be high intensity and fast spreading with high fire temperatures, the Sycan Marsh burn to be slower spreading with lower fire temperatures (some high temperature hotspots), and the Shoetank Rx burn to be slower spreading with lower fire temperatures.

\begin{figure}[t]
    \centering
    \includegraphics[width=\columnwidth]{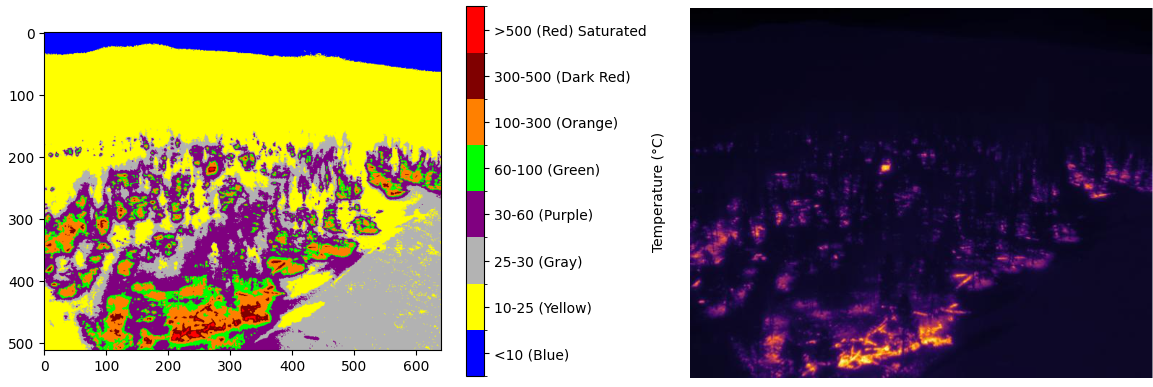}
    \caption{\centering TIFF Temperature Range Visualization - Sycan Marsh Sample; Temperature Graph (Left), Thermal JPG (Right)}
    \label{fig:tiffcalibration_expectedtemps}
\end{figure}

The saturation point, or temperature at which the thermal camera becomes unreliable, of the M2EA and M30T thermal cameras used in the FLAME 3 data collection, in low gain mode are 450 and 500 degrees Celsius, respectively. The TIFFs in FLAME 3 were analyzed quantitatively and visually to understand if pixels were saturating, as seen in  {Figure \ref{fig:tiffcalibration_expectedtemps}}.

From visual observation, all temperature values were accurate, with the exception of a low quanity of saturated pixels. In  {Figure \ref{fig:tiffcalibration_expectedtemps}}, the blue region of temperatures less than 10 degrees Celsius is representative of the sky. Additionally, it was observed that the main fire regions across all burns fell between 100 and 450-500 degrees Celsius. We decided to clip the TIFF data values between 0 and 500 degrees Celsius. This effectively removed the negative sky temperatures to create a more consistent temperature for the non-fire region and gets rid of unreliable temperature readings above the M30T saturation point. If the M2EA saturation point of 450 degrees Celsius was chosen as the value to clip at, then this would affect 450-500 reliable temperature readings in the Sycan Marsh data. Saturated pixels make up a small portion of the data, but any temperature predictions between 450-500 Celsius should be considered carefully.

\section{Related Work}

AI-based wildfire perception models have gained significant traction in recent years, particularly with the availability of UAV-collected datasets. These datasets have facilitated the development of vision models for tasks including flame segmentation, smoke detection, and fire line tracking. Early models primarily focused on unimodal segmentation using RGB imagery due to its wide availability and low cost. Architectures such as DeepLabV3+ \cite{deeplabv3+} and ResNet-based U-Nets have been widely adopted for wildfire segmentation tasks. While effective under ideal conditions, RGB-only models are limited in their ability to detect flames under occlusion (e.g., smoke), in low-light settings, or in visually ambiguous regions where thermal information is crucial \cite{flamefinder}.

 Multimodal architectures that fuse RGB and thermal inputs have shown promising performance for semantic segmentation. These include MFNet \cite{mfnet}, RTFNet \cite{rtfnet}, FEANet \cite{feanet}, and SFAF-MA \cite{SFAFMA}, which use dual-encoder structures and cross-modality fusion blocks to combine complementary features from each modality. In the wildfire context, models such as FlameFinder \cite{rajoli2024} and works leveraging FLAME 2 imagery \cite{wildfiresegmentation1} have shown that thermal cues can significantly enhance segmentation accuracy. However, these methods require thermal sensors at inference time, which increases cost, weight, and power consumption—constraints that are prohibitive for low-SWaP (Size, Weight, and Power) UAV platforms. This motivates the need for models that can learn from multimodal data but deploy with unimodal inputs.

Recent advances in knowledge distillation offer a promising pathway to this goal. In modality distillation, a multimodal teacher model supervises the training of a unimodal student, transferring its knowledge without requiring access to privileged modalities during inference. This strategy allows lightweight models to approximate the performance of larger, sensor-rich systems. While knowledge distillation has been applied in other domains, its use in wildfire perception remains limited. A notable effort by Pesonen et al. \cite{uavteacher} used SAM-generated pseudo labels to distill smoke segmentation knowledge from an RGB teacher to an RGB student network, demonstrating that distillation can reduce the need for annotated data and simplify inference. However, their model was limited to binary smoke segmentation and did not incorporate temperature inference or radiometric ground truth.

To the best of our knowledge, our work is the first to extend this framework to the dual tasks of wildfire segmentation and per-pixel temperature regression. By leveraging radiometric thermal TIFFs from FLAME 3 as supervision and combining foundation models like SAM with traditional CV methods to generate high-quality pseudo labels, we propose a new training paradigm that unifies multimodal learning and knowledge distillation. This approach enables robust wildfire monitoring using only RGB inputs, significantly improving deployment feasibility without sacrificing the benefits of thermal sensing.

\begin{figure*}[t]
    \centering
    \includegraphics[width=0.95\textwidth, height=9 cm, trim=0 0 0 0, clip]{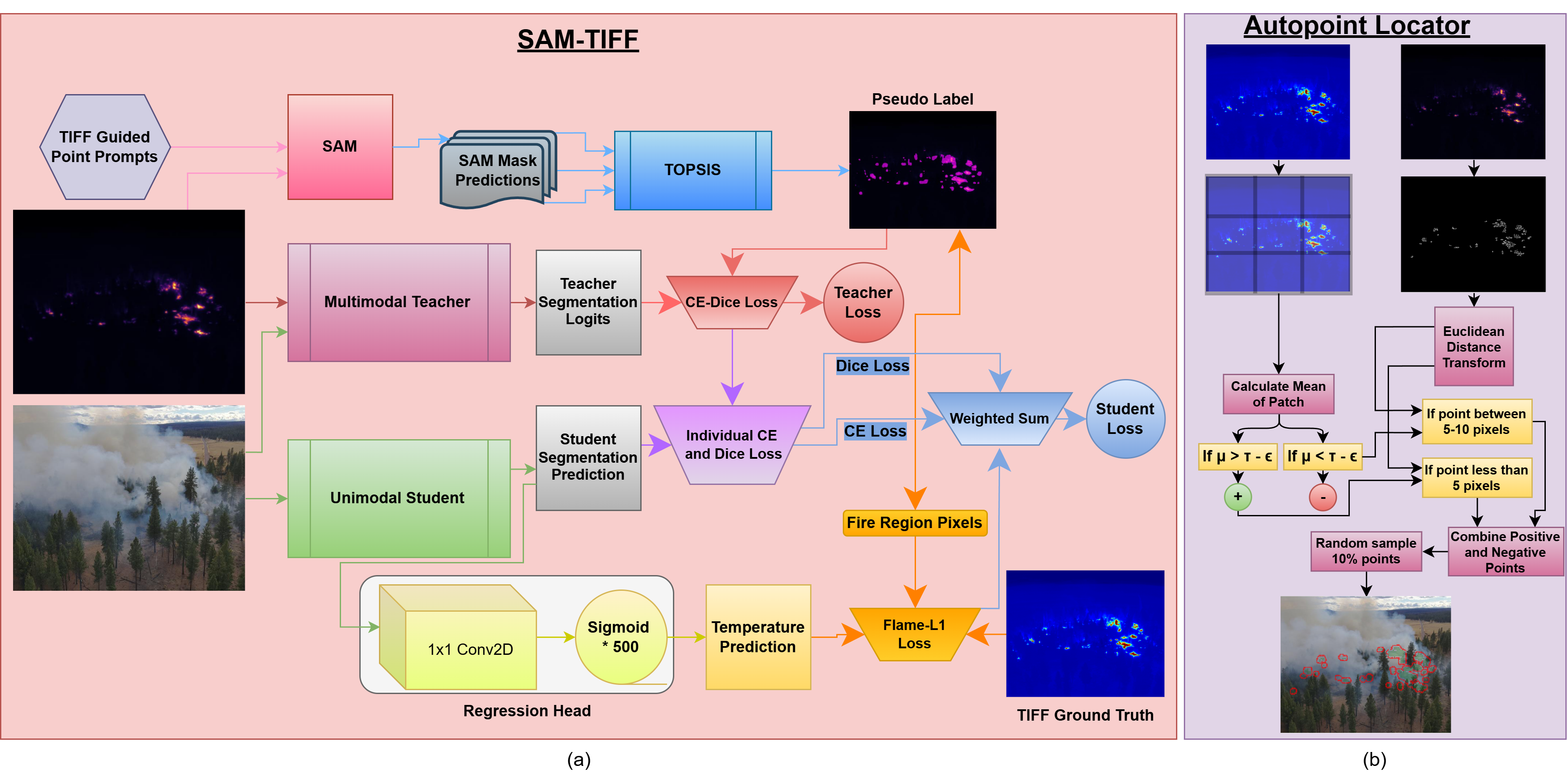}
    \caption{\centering SAM-TIFF Architecture (a), Autpoint Locator Algorithm (b) - `TIFF Guided Point Prompts' Module in (a); Jet Colormap for TIFF Visualization, Inferno Colormap for Thermal JPG Visualization; Canny Edge Detection Thresholds - Low Threshold = \(\tau\) and High Threshold = 200 degrees Celsius; $\epsilon$ = 25 degrees Celsius
}
    \label{fig:finalarch}
\end{figure*}

\section{SAM-TIFF Temperature Inference}

In this section, we describe the SAM-TIFF  framework, where a multimodal (RGB-T) teacher network distills knowledge to a unimodal (RGB) student network for both fire segmentation and pixel level temperature prediction using the FLAME 3 Dataset \cite{flame_3_dataset}. The core idea around this work is that if the RGB input only student network could learn from the more informed RGB-T teacher network, the result would be a standalone student capable of predicting per-pixel temperature from an RGB image only. This method could render the use of IR sensors and thermal cameras in UAV systems unnecessary. Additionally, this work could enable devices, such as smartphones, without thermal cameras to be used in wildfire or other temperature-centered scenarios for hotspot detection.

As shown in Fig. \ref{fig:finalarch}a, the proposed SAM-TIFF architecture consists of two stages: (1) a multimodal RGB-Thermal teacher model fine-tuned on FLAME 3 with SAM-generated segmentation mask pseudo labels, and (2) an RGB-only student model trained to replicate the segmentation outputs of the teacher and predict temperature using pseudo label fire region guidance.  The multimodal teacher network is pretrained as described in Sec. \ref{sec:flame2_pretraining}.  The temperature head of the student network uses a regression layer with sigmoid scaling to constrain outputs to the physical temperature range captured in FLAME 3 (0–500°C). A composite loss function jointly optimizes segmentation and temperature accuracy, with segmentation guided by cross-entropy and Dice losses, and temperature supervised using a region-masked L1 loss that focuses training only on pixels identified as fire by the SAM pseudo labels.

There are five main loss functions we compute during each training epoch between, the teacher and SAM pseudo labels ($\mathcal{L}_{\text{teacher}}$, {Equation \ref{eq:finalteacherloss}}), the student and teacher segmentation prediction ($\mathcal{L}_{\text{student\_cross}}$ and $\mathcal{L}_{\text{student\_dice}}$), the student temperature prediction and TIFF ground truth ($\mathcal{L}_{\text{Flame-L1}}$), and a total student loss ($\text{L}_{\text{student}}$). Two instances of Adam optimizer are used, one for the teacher and one for the student. 

\begin{enumerate}[leftmargin=*]
    \item \begin{equation}
\mathcal{L}_{\text{teacher}} = \mathcal{L}_{\text{CE}} + \lambda_{\text{Dice}} \cdot \mathcal{L}_{\text{Dice}}
\label{eq:finalteacherloss}
\end{equation}
where, \( \mathcal{L}_{\text{CE}} \) is the cross-entropy loss between the teacher and SAM pseudo label, \( \mathcal{L}_{\text{Dice}} \) is the dice loss between the teacher and SAM pseudo label, and \( \lambda_{\text{Dice}} \) is a hyperparameter controlling the relative importance of \( \mathcal{L}_{\text{Dice}} \). The \( \lambda_{\text{Dice}} \) term was set to 0.5 for this study.

\item $\mathcal{L}_{\text{student\_cross}}$ and $\mathcal{L}_{\text{student\_dice}}$ are individually computed cross entropy loss and dice loss between the student segmentation prediction and the teacher segmentation prediction, respectively.

\item Fire-region temperature regression loss is computed only over fire region pixel locations from the SAM pseudo label as: \begin{equation}
\mathcal{L}_{\text{Flame-L1}} = 
\begin{cases}
\frac{1}{N}\sum_{i \in \mathcal{F}} \left| \hat{T}_i - T_i \right|, & \text{if } N > 0 \\
0, & \text{if } N = 0
\end{cases}
\label{eq:flamel1loss}
\end{equation}
$\hat{T}_i$ is the predicted temperature at pixel $i$ from the student model,  $T_i$ is the ground truth temperature at pixel $i$ from thermal TIFF, $\mathcal{F}$ denotes the set of pixel indices corresponding to fire regions, defined as $\mathcal{F} = \{ i \mid SAM_i = 1 \}$, where $SAM_i$ is the SAM-generated fire mask,  $N = |\mathcal{F}|$: Number of pixels in the fire region. If $N = 0$, the loss is set to zero, and $\left| \hat{T}_i - T_i \right|$: Absolute difference between the predicted and ground truth temperature at pixel $i$ (L1 loss portion).

\item 
\begin{equation}
\resizebox{0.95\linewidth}{!}{$
\text{L}_{\text{student}} = \text{L}_{\text{student\_cross}} + \lambda_{\text{student\_dice}} \cdot \text{L}_{\text{student\_dice}} + \lambda_{\text{Flame-L1}} \cdot \text{L}_{\text{Flame-L1}}
$}
\label{eq:totalflameloss}
\end{equation}
where,  $\lambda_{\text{student\_dice}}$ is the weight for $\text{L}_{\text{student\_dice}}$ term, and 
 $\lambda_{\text{Flame-L1}}$ denotes the weight for $\text{L}_{\text{Flame-L1}}$ term.
\end{enumerate}

During each training epoch, the teacher network is first updated while keeping the student network parameters frozen. After the teacher optimization step, the student network is trained while the teacher network parameters are frozen. To ensure correct gradient flow, teacher predictions used in the student loss calculations are detached from the computation graph. This process guarantees that no parameter updates are made to the teacher during student training and that loss calculations for the student do not affect the teacher's weights.

\section{Hybrid Mask Generation }
\label{sec:hybridcv}
FLAME 3 \cite{flame_3_dataset} does not provide ground truth for segmentation. Since this paper aims to design an architecture that can segment fire and predict temperature, segmentation ground truth masks need to be created. We design a pipeline that incorporates TIFF temperature data with Canny Edge Detection and Otsu's Method to identify positive and negative point prompts for SAM. These point prompts along with the thermal JPG are input to the SAM Predictor \cite{sam}, which outputs three masks and associated confidence scores. Finally, rather than using the raw SAM confidence score, where the highest confidence score does not always align with the desired mask prediction, we define five criteria used with the multi-criteria decision-making method TOPSIS \cite{TOPSIS} to choose the final ground truth mask.

\subsection{Positive and Negative Point Autolocator}
The procedure for locating positive and negative points, and filtering out unwanted points using Canny Edge Detection and a Euclidean Distance Transform, is shown in Fig. \ref{fig:finalarch}b. The patch size used for positive points was 5x5 and for negative points, was 3x3. In this figure, $\mu$ and $\tau$ denote the TIFF patch mean and the optimal threshold computed per image from Otsu, respectively.  $\epsilon$ is the error margin applied to counteract Otsu foreground misclassification.

After the positive and negative points are located, it was observed that for the FLAME 3 images, SAM shows worse performance if there are too many positive and negative points in unnecessary regions. Therefore, we aim to filter out unwanted points and center them around fire edges identified with Canny Edge Detection. The Euclidean Distance Transform contains the shortest distance of each point to the nearest Canny edge pixel. The procedure for this filtering process is also shown in Fig. \ref{fig:finalarch}b.

\subsection{Final Ground Truth Mask Decision}
Manually selecting the optimal mask from the three outputs generated by SAM Predictor is both labor-intensive and inconsistent. To automate this process, we employ the Technique for Order Preference by Similarity to Ideal Solution (TOPSIS), a multi-criteria decision-making framework. This approach enables objective mask selection by evaluating each candidate based on five quantitative criteria designed to capture foreground accuracy and structural similarity. The criteria used in the TOPSIS evaluation are:

\begin{itemize}[leftmargin=*]
    \item IoU (Otsu vs. SAM): Measures the Intersection over Union between the SAM mask and the mask generated via Otsu’s method. This criterion favors masks that closely resemble the Otsu-derived segmentation, which serves as a classical baseline.

\item IoU (Thresholded Thermal JPG vs. SAM): Quantifies the overlap between the SAM mask and a thresholded thermal image. Higher values indicate better alignment with thermal features and are thus prioritized.

\item Mean Foreground Temperature Difference: Computes the absolute mean temperature difference between foreground pixels in the SAM mask and those in the Otsu mask using radiometric TIFF data. A lower difference suggests the mask is capturing a similar thermal foreground region as Otsu and is therefore preferred.

\item SAM Confidence Score: Incorporates the original confidence score provided by SAM. Although this is not always indicative of the best segmentation, it is included as a supplementary metric.

\item Structural Similarity Index (SSIM - Otsu vs. SAM): Evaluates the perceptual similarity in structure and edges between the SAM mask and the Otsu mask. This encourages the preservation of coherent object boundaries.
\end{itemize}
To reflect the higher reliability of thermal-based cues, the IoU between the thresholded thermal JPG and SAM is given greater weight in the TOPSIS scoring. The final mask selected is the one with the highest TOPSIS score across all criteria.

Following the automated selection, all generated masks in the FLAME 3 computer vision (CV) subset were manually reviewed. A small number of masks that did not meet quality expectations were removed from the dataset used in subsequent experiments. The final image counts per burn location are summarized in Table \ref{tab:cvfinalimagecounts}.

\begin{table}[h]
    \centering
    \caption{\centering  {\small FLAME 3 CV Subset - Final Image Pair (RGB-Thermal-SAM Mask-TIFF) Counts Used by Burn Location}}
    \begin{tabular}{ccc}  
        \toprule
        Burn Location & Excluded Count & Final Count \\
        \midrule
        Shoetank & 554 & 731 \\
        Sycan2A & 40 & 324 \\
        Sycan2D & 33 & 225 \\
        Willamette Valley & 0 & 232 \\
        \midrule
        All Burn Locations & 627 & 1512 \\
        \bottomrule
    \end{tabular}
    \vspace{-10pt}
    \label{tab:cvfinalimagecounts}
\end{table}

\section{FLAME 2 Pretraining}
\label{sec:flame2_pretraining}
Training the teacher network from scratch alongside the student network, could result inhibit student learning in early epochs. Additionally, using a teacher pre-trained on ImageNet is not an invalid approach, but ideally, we want a teacher with some prior wildfire domain knowledge before the final design training. To avoid these pitfalls, it made sense to use FLAME 2 \cite{flame_2_dataset} and pretrain various multimodal networks to be used as a teacher network in the final design. FLAME 2 only includes labels for Fire / No Fire and Smoke / No Smoke, so segmentation ground truth needed to be created. All fire images were extracted from FLAME 2 and Otsu's Method \cite{otsu} was used with the three channel thermal JPG (converted to greyscale for Otsu) as input. This resulted in 39,751 image pairs with ground truth Otsu Masks. 

Once the ground truth was created, various multimodal encoder-decoder networks were trained on a subset of FLAME 2 for wildfire binary segmentation \cite{hopkins2}. The experimental setup is as follows. Six networks were trained separately: FEANet \cite{feanet}, RTFNet \cite{rtfnet}, MFNet \cite{mfnet}, U-Net \cite{unet}, EAEFNet \cite{eaef}, and SFAFMA \cite{SFAFMA}. These networks and different ResNet backbones were used and the pretrained ImageNet1K V2 weights (U-Net used ImageNet1K V1) were used, with the exception of MFNet which does not use a ResNet backbone.

Pytorch's Exponential Learning Rate Scheduler was used with an initial learning rate of 0.001 and a decay rate of 5\% each epoch. Adam optimizer was used and the weight decay was initially set to 0.00005 and was exponentially decayed by 5\% each epoch. Due to the large size and high temporal resolution of FLAME 2, a subset was needed to prevent overfitting. The full dataset train-validation-test split was 80-10-10 and then 10000 training, 1000 validation, and 1000 testing images were randomly sampled from this. A seed was set for each network training to ensure the same split was taken each time and that results could be compared. The batch size was set to 16 and each network was trained for 101 epochs, with validation after each epoch. To further prevent overfitting, the transforms applied to each RGB and Thermal images individually were resizing from original resolution (2160h x 3840w) to a smaller resolution (512h x 640w), converting to PyTorch tensors, applying a min-max normalization (to ensure input tensors were on the same scale), Color Jitter, and Gaussian Blur. We adopt the loss function from {Equation \ref{eq:finalteacherloss}} for this pretraining stage.

In {Table \ref{tab:pretraining_testing}}, the values recorded are from epoch 101, with the exception of the EAEF \cite{eaef} networks, which are from epoch 21, as this network converged more quickly than the others. The mAcc and mIoU reported are the averages across all testing batch averages. We calculate the number of correctly classified pixels for both the foreground and background classes individually. The mAcc is the average between these two values.

In  {Table \ref{tab:pretraining_testing}}, mIou and mAcc are reported for each multimodal network and backbone. The highest performing network overall was SFAFMA-50, achieving 96.69\% mIoU and 98.74\% mAcc. These metrics indicate balanced performance across both foreground and background classes. It is important to note that these results indicate some degree of overfitting, most likely caused by the ground truth being generated from Otsu's Method directly from the thermal JPGs themselves. This lends itself to these networks potentially learning the thresholding of Otsu's Method or thermal colormapping during training rather than the true segmentation of the fire. A more robust ground truth creation technique could potentially avoid this, but is out of the scope of this work. For purposes of this work, the results are satisfactory for the pretraining phase. Checkpoints for each network were saved and the respective weights loaded as teacher weights during the modality distillation phase. 
\begin{table}[t]
\footnotesize
\centering
\caption{{\small FLAME 2 Pretraining – Overall mIoU and Accuracy for Various Models and Backbones}}
\label{tab:pretraining_testing}
\begin{tabular}{l l c c}
    \toprule
    \textbf{Model} & \textbf{Backbone} & \textbf{mIoU} & \textbf{Accuracy} \\
    \midrule
    MFNet \cite{mfnet}       & None         & 0.9563 & 0.9841 \\
    UNet \cite{unet}         & ResNet-50    & 0.9623 & 0.9846 \\
    UNet \cite{unet}         & ResNet-152   & 0.9610 & 0.9861 \\
    FEANet \cite{feanet}     & ResNet-50    & 0.9424 & 0.9764 \\
    FEANet \cite{feanet}     & ResNet-101   & 0.9454 & 0.9799 \\
    FEANet \cite{feanet}     & ResNet-152   & 0.9390 & 0.9762 \\
    RTFNet \cite{rtfnet}     & ResNet-101   & 0.9377 & 0.9757 \\
    RTFNet \cite{rtfnet}     & ResNet-152   & 0.9368 & 0.9753 \\
    EAEFNet \cite{eaef}      & ResNet-50    & 0.9409 & 0.9792 \\
    EAEFNet \cite{eaef}      & ResNet-152   & 0.9471 & 0.9716 \\
    SFAFMA \cite{SFAFMA}     & ResNet-50    & \textbf{0.9669} & \textbf{0.9874} \\
    \bottomrule
\end{tabular}
\vspace{-5pt}
\end{table}

\section{Results} 
For training and evaluation, teacher networks pretrained on FLAME 2 \cite{flame_2_dataset} from {Section \ref{sec:flame2_pretraining}} were used with a student network U-Net \cite{unet} (ResNet-152 backbone) or DeepLabV3+ \cite{deeplabv3+} (ResNet-101 backbone). The transformations applied to the RGB images were converting to a PyTorch tensor, Min-Max normalization between range (0-1), Color Jitter, Random Gamma, Random Gaussian Noise, and Gaussian Blur. The transformations applied to the thermal image were converting to a PyTorch tensor, Min-Max normalization between the range (0-1), and Gaussian Blur. All CV subset locations and the final image pair counts, listed in  {Table \ref{tab:cvfinalimagecounts}}, were used. An 80-20 train-test split was used, with no validation due to the low CV subset image count available. Adam optimizer was used for the teacher and student optimizer, with a learning rate of 1e-4 for the teacher and 1e-3 for the student. The teacher learning rate was purposely set lower for fine-tuning and as it was pretrained on FLAME 2 and already had some level of learned wildfire segmentation knowledge. The weight decays were 1e-5 for the teacher and 1e-4 for the student. The networks were trained for 270 epochs and no learning rate scheduler was used. EAEF-50 converged faster than most networks, so epoch 150 is reported in the results tables. For evaluation of our architecture, we report IoU per class, mIoU, and per-pixel accuracy per temperature tolerance range. The pixel accuracy measures the percentage of correctly predicted temperature pixels that are within plus or minus of the corresponding ground truth TIFF temperature pixels. The mean metrics are representative of the average of each batch mean for the whole test set. 

\begin{table}[t]
    \centering
    \caption{\centering Teacher-Student Combination Testing IoU per Class and mIoU; Class 0 = Background, Class 1 = Fire Region}
    \label{tab:teacherstudent_miou}
    \resizebox{0.48\textwidth}{!}{ 
    \begin{tabular}{l c c c c c} 
        \toprule
        \textbf{Teacher} & Teacher Backbone & \textbf{Student} & \textbf{IoU 0} & \textbf{IoU 1} & \textbf{mIoU} \\
        \midrule
        MFNet \cite{mfnet} & None     & U-Net & 0.9515 & 0.4403 & 0.6959 \\
        UNet \cite{unet}   & ResNet-50    & U-Net & 0.9503 & 0.4451 & 0.6977 \\
        UNet \cite{unet}   & ResNet-152    & U-Net & 0.9497 & 0.4423 & 0.6960 \\
        FEANet \cite{feanet} & ResNet-50  & U-Net & 0.9498 & 0.4352 & 0.6925 \\
        FEANet \cite{feanet} & ResNet-101  & U-Net & 0.9531 & 0.4213 & 0.6872 \\
        FEANet \cite{feanet} & ResNet-152  & U-Net & 0.9461 & 0.4452 & 0.6956 \\
        RTFNet \cite{rtfnet} & ResNet-101  & U-Net & 0.9569 & 0.4230 & 0.6900 \\
        RTFNet \cite{rtfnet} & ResNet-152 & U-Net & 0.9549 & 0.4649 & 0.7099 \\
        EAEFNet \cite{eaef}  & ResNet-50  & U-Net & 0.9485 & 0.4079 & 0.6782 \\
        EAEFNet \cite{eaef}  & ResNet-152  & U-Net & 0.9497 & 0.4286 & 0.6891 \\
        SFAFMA \cite{SFAFMA} & ResNet-50  & U-Net & 0.9517 & 0.4579 & 0.7048 \\
        \midrule
        MFNet \cite{mfnet}   & None        & DeepLabV3+    & 0.9595 & 0.4466 & 0.7031 \\
        FEANet \cite{feanet}  & ResNet-50  & DeepLabV3+    & 0.9559 & 0.4538 & 0.7048 \\
        FEANet  \cite{feanet} & ResNet-101 &  DeepLabV3+   & 0.9581 & 0.4531 & 0.7056 \\
        FEANet \cite{feanet}  & ResNet-152 &  DeepLabV3+    & 0.9543 & 0.4243 & 0.6893 \\
        RTFNet \cite{rtfnet}   & ResNet-101 &  DeepLabV3+   & \textbf{0.9637} & 0.4355 & 0.6996 \\
        EAEFNet \cite{eaef}  & ResNet-50  &  DeepLabV3+    & 0.954 & 0.4398 & 0.6969 \\
        EAEFNet \cite{eaef} & ResNet-152 & DeepLabV3+    & 0.9594 & 0.4413 & 0.7004 \\
        SFAFMA \cite{SFAFMA} & ResNet-50 &  DeepLabV3+   & 0.9561 & \textbf{0.4728} & \textbf{0.7144} \\
        \bottomrule
    \end{tabular}
    }
\end{table}

\begin{figure}[t]
    \centering
    \includegraphics[width=0.45\textwidth]{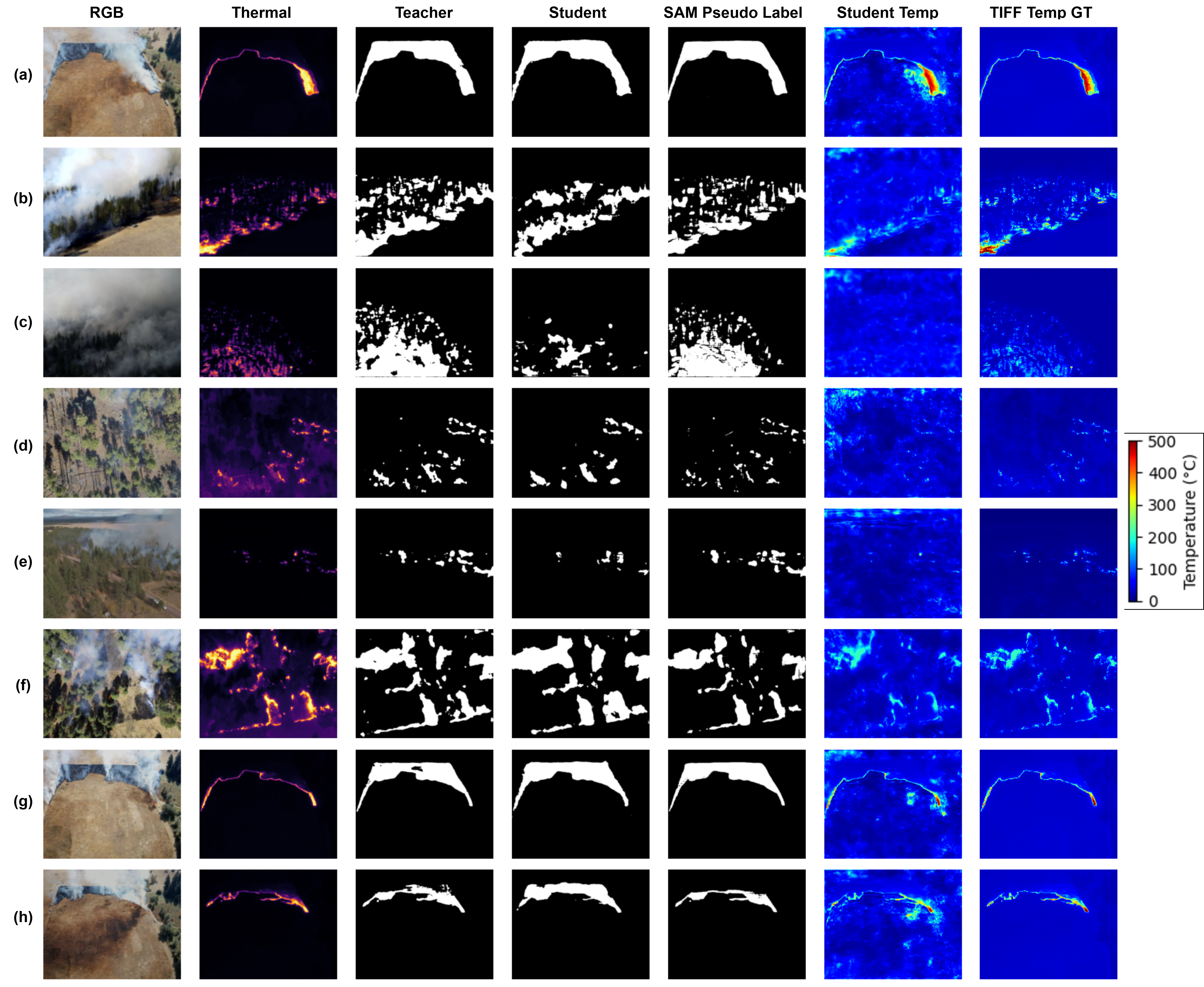}
    \caption{\centering Segmentation predictions for SFAFMA-50 Teacher - DeepLabV3+ Student Variant}
    \label{fig:studentnetworkproblems}
\end{figure}

 {Table \ref{tab:teacherstudent_miou}} shows the pretrained teacher networks, teacher backbone, and the U-Net or DeepLabV3+ student network, testing results. It is seen that the highest mIoU that was achieved was a DeepLabV3+ student learning from SFAFMA-50 teacher, achieving a 71.44\% mIoU. SFAFMA-50 also showcased the best performance in fire region class IoU, 47.28\% accuracy. The DeepLabV3+ student showcased better performance overall with a smaller ResNet-101 than the U-Net ResNet-152 student. At first glance, the IoU for the fire regions may seem low, but after observing visual results, it became clear that certain burn imagery was difficult for the student network to learn. This meant that the student network was predicting highly accurate for some burn locations, but not as accurate for others. Some images in burns such as Willamette Valley are more consistent and have a higher temporal resolution than the Sycan Marsh burn. Additionally, some imagery in FLAME 3 contains views of smoke and trees only, and no visible fire in the image. With a three-channel RGB color image only as input, and no distinct fire colors in the image, it may have proven difficult for the student network to segment the fire region. Some of these difficulties are visualized in  {Figure \ref{fig:studentnetworkproblems}}, rows b - e, reflecting not necessarily poor, but not ideal results. In summary, the overall sporadic nature and no visible flames of some of the burn imagery most likely caused lower quantitative IoU for the fire region (Class 1). 

\begin{figure}[t]
    \centering
    \includegraphics[width=0.45\textwidth]{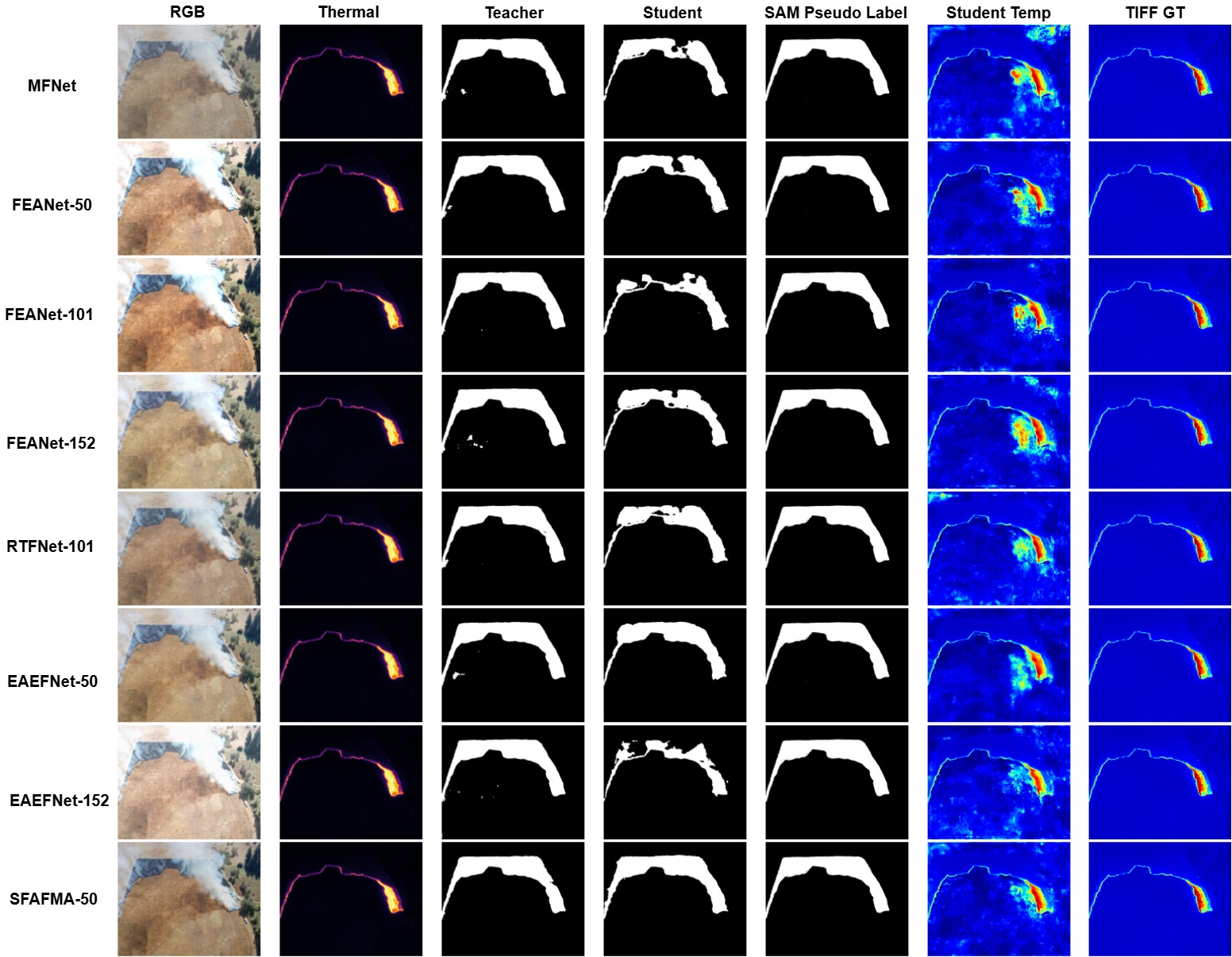}
    \caption{\centering Willamette Valley Sample Image Results for Teacher Networks (left) - DeepLabV3+ Student Variants - Temp Predictions [Jet Colormap] between 0 and 500 (\(^\circ\)C)}
    \label{fig:test1sample}
\end{figure}

Sample visual results for a test image from Willamette Valley for the teachers with DeepLabV3+ student network are shown in  {Figure \ref{fig:test1sample}}.

\begin{table}[t]
    \centering
    \caption{\centering Fire Region Pixel Percentage Accuracy - Within $\pm$ Tolerance (\(^\circ\)C)}
    \label{tab:teacherstudent_acc}
    \resizebox{0.48\textwidth}{!}{ 
    \begin{tabular}{l c c c c c} 
        \toprule
        \textbf{Teacher} & Teacher Backbone & Student &\textbf{$\pm$25} & \textbf{$\pm$50} \\
        \midrule
            MFNet \cite{mfnet} & None & U-Net & 67.56 & 83.13 \\
            UNet \cite{unet} & ResNet-50 & U-Net & 67.08 & 83.35 \\
            UNet \cite{unet}   & ResNet-152 & U-Net &  66.38 & 82.72 \\
            FEANet \cite{feanet} & ResNet-50 & U-Net & 67.84 & 83.18 \\
            FEANet \cite{feanet} & ResNet-101 & U-Net & 67.21 & 82.98 \\
            FEANet \cite{feanet} & ResNet-152 & U-Net & 67.45 & 83.34 \\
            RTFNet \cite{rtfnet} & ResNet-101 & U-Net & 68.04 & 83.60 \\
            RTFNet \cite{rtfnet} & ResNet-152 & U-Net & 68.39 & 83.49 \\
            EAEFNet \cite{eaef}  & ResNet-50 & U-Net & 67.00 & 83.50 \\
            EAEFNet \cite{eaef} & ResNet-152 & U-Net & 67.19 & 82.88 \\
            SFAFMA \cite{SFAFMA} & ResNet-50 & U-Net & 67.82 & 83.30 \\
        \midrule
            MFNet \cite{mfnet}   & None        & DeepLabV3+ & 68.62 & 84.30 \\
            FEANet \cite{feanet}  & ResNet-50 & DeepLabV3+ & 68.80 & 84.98 \\
            FEANet \cite{feanet} & ResNet-101 & DeepLabV3+ & 68.72 & 84.45 \\
            FEANet \cite{feanet}  & ResNet-152 & DeepLabV3+ & 67.95 & 83.92 \\
            RTFNet \cite{rtfnet}   & ResNet-101 & DeepLabV3+ & 68.48 & 84.37 \\
            EAEFNet \cite{eaef}  & ResNet-50  &  DeepLabV3+ & 68.61 & 84.63 \\
            EAEFNet \cite{eaef} & ResNet-152 &  DeepLabV3+ & \textbf{68.95} & \textbf{84.83} \\
            SFAFMA \cite{SFAFMA} & ResNet-50 & DeepLabV3+ & 68.78 & 84.68 \\
        \bottomrule
    \end{tabular}
    }
\end{table}

 {Table \ref{tab:teacherstudent_acc}} shows testing results with different teacher-student variants of the temperature predictions for the ground truth fire region pixels only. These reflect the output from the student regression head modification. The percentage reported is the percentage of predicted pixels that are within plus or minus the given tolerance value. For the fire region pixels, the SAM pseudo label fire region pixel locations were extracted and then those locations were used to compare the predicted student network temperatures to the ground truth TIFF temperatures. It is seen that for fire region temperature prediction, all student networks achieved close to 68\% accuracy within $\pm$25 degrees Celsius and around 83-84\% accuracy within $\pm$50 degrees Celsius. EAFFNet and SFAFMA teachers with DeepLabV3+ student variants provide the best performance. High end drone IR sensors can achieve accuracy within $\pm$2 degrees Celsius, so the results shown are not quite at that level of precision yet. The \(Flame_{L1}\) loss was designed to focus on fire region temperature prediction only, and future improvements may yield better results. Also, the \(Flame_{L1}\) loss was included as a singular term in the total weighted student loss, \(L_{student}\), which included consideration for segmentation accuracy as well. The benefit to using \(L_{student}\) is that the two tasks, segmentation and temperature prediction, are optimized at the same time. However, the downside is that one task may dominate the loss or struggle to learn at the same rate as the other. Overall, SFAFMA-50 as the teacher and a DeepLabV3+ student provides the best balance between segmentation accuracy and temperature accuracy.

\section{Conclusions and Future Work}
By decoupling the training and inference sensor requirements, the proposed SAM-TIFF opens the door to scalable, low-cost wildfire perception systems with the accuracy of multimodal sensing and the deployability of RGB-only UAVs. Our approach lays the foundation for next-generation aerial monitoring systems that combine the power of foundation models, multimodal learning, and distillation to deliver mission-critical intelligence in complex environmental conditions.

\bibliographystyle{IEEEtran}
\bibliography{references}
\end{document}